\documentclass[slac_one]{revtex4}
\usepackage{graphicx}
\usepackage{fancyhdr}
\begin{document}

\title{Charmed $B(B_{s})$ decays involving a light tensor meson in PQCD approach} 

%

\author{Zhi-Tian Zou}
\author{Xin Yu}
\author{Cai-Dian L\"{u}}
\affiliation{Institute  of  High  Energy  Physics  and  Theoretical  Physics Center for Science Facilities,
Chinese Academy of Sciences, Beijing 100049, People's Republic of China }

\begin{abstract}
We study the $B(B_{s})\rightarrow D_{(s)}(\bar{D}_{(s)})\,T$ and $D_{(s)}^{\ast}(\bar{D}_{(s)}^{*})\,T$ decays in perturbative QCD approach, where T denotes a light tensor meson. In addition to the branching ratios, we also give predictions to the polarization fractions for the $D^{*}_{(s)}(\bar{D}_{(s)}^{*})T$ decays. For those decays with a tensor meson emitted, the factorizable emission diagrams do not contribute because of lorentz covariance considerations. In order to give the predictions to these decay channels, it is necessary to go beyond the naive factorization to calculate the nonfactorizable  and annihilation diagrams.
\end{abstract}

\maketitle

\thispagestyle{fancy}


\section{INTRODUCTION} 
Recently, the measurements about B decays involving a light tensor meson (T) have been
 making progress \cite{jpg37075021}.
  For the tensor meson considered in this work, $J^{p}=2^+$, the quark pair's orbital
 angular momentum L and total spin S are equal
 to 1. However, their production property in B decays is
 quite similar to the light vector mesons, because the polarizations with $\lambda=\pm2$ do not
 contribute due to the angular momentum conservation \cite{wwprd83014008}.
These rare B decays have been studied in the naive factorization
\cite{epjc22683epjc22695,prd67014002}. In this work, we consider
these B decays, which have been discussed in the factorization
approach \cite{prd491645}, involving a heavy $D$ meson and a tensor
meson in the final states. Due to the fact that $\langle 0\mid
j^{\mu}\mid T \rangle = 0$, where $j^{\mu}$ is the $(V\pm A)$ or
$(S\pm P)$ current \cite{zheng1zheng2,epjc22683epjc22695}, there is
no contribution form the factorizable diagrams with a tensor meson
emitted. Because of this difficulty, the naive factorization
approach can not give the right prediction. So we must go beyond the
naive factorization to calculate the nonfactorizable contributions
and the annihilation type contributions.  In this paper, we use the perturbative QCD factorization approach
(PQCD) \cite{zou} to calculate these contributions

There are various energy scales involved in hadronic $B(B_s)$
decays, i.e., three different scales: W
boson mass scale, B meson mass scale $M_{B}$ and the factorization
scale $\sqrt{\bar{\Lambda}M_{B}}$, where $\bar{\Lambda}\equiv
M_{B}-m_{b}$. The electroweak physics higher than W boson mass can
be calculated perturbatively. By using the renormalization group
equation, the physics between B meson mass scale and W boson mass
scale can be included in the Wilson coefficients of the effective
 four-quark operators. The hard part in the PQCD approach includes the physics between $M_{B}$ and the factorization scale.
 The physics below the factorization scale is nonperturbative, which can be described by the universal hadronic wave
 functions of mesons. In PQCD approach, in order to avoid the endpoint singularity which
 spoil the perturbative calculation, we do not neglect the transverse momentum $k_{T}$ of the light quarks in meson. As a result,
 the additional energy scale will appear, which is introduced by the transverse momentum. The additional scale can give large double
 logarithms in the hard part calculations.
 We can resum these double logarithms by using the renormalization group equation to give a Sudakov
factor, which effectively suppresses the end-point region
contribution and makes the PQCD approach more reliable and
consistent.

\section{FORMALISM}
We know that the light quark in $B(B_{s})$ meson is soft, because the heavy b quark carries most of the energy of $B(B_{s})$
 meson. But it is collinear in the final state light
meson, so a hard gluon is necessary to connect the spectator quark
to the four quark operator. The hard part of the interaction in PQCD
approach contains six quarks rather than four quarks, which is
called six-quark effective theory or six-quark operator. The decay
amplitude can be explicitly factorized as the following formalism:
\begin{eqnarray}
\mathcal
{A}\;\sim\;&&\int\,d[x]b_{1}db_{1}b_{2}db_{2}b_{3}db_{3}\times
Tr\left[C(t)\Phi_{B}(x_{1},b_{1})\Phi_{M_{2}}(x_{2},b_{2})\Phi_{M_{3}}(x_{3},b_{3})H(x_{i},b_{i},t)S_{t}(x_{i})e^{-S(t)}\right],
\end{eqnarray}
where $b_{i}$ is the conjugate variable of transverse momentum
$k_{iT}$. $x_{i}$ is the momentum fractions in mesons.  $t$ is the
largest energy scale in calculation of $H(x_{i},b_{i},t)$. $C(t)$
are the corresponding Wilson coefficients of four quark operators.
$\Phi(x)$ are the meson wave functions. $S_{t}(x_{i})$ is obtained
by the threshold resummation, that smears the end-point singularities
on $x_{i}$. The $e^{-S(t)}$ term is the so-called Sudakov form
factor. It can effectively suppresses the soft dynamics and the long
distance contributions in the large $b$ region.

The polarization tensor  $\epsilon_{\mu\nu}(\lambda)$ with helicity
$\lambda$ of tensor meson can be expanded via the vectors
$\epsilon^{\mu}(0)$ and $\epsilon^{\mu}(\pm1)$, which are the
polarization vector of vector meson, \cite{zheng1zheng2}
\begin{eqnarray}
\epsilon^{\mu\nu}(\pm2)\,&\equiv&\,\epsilon(\pm1)^{\mu}\epsilon(\pm1)^{\nu},\nonumber\\
\epsilon^{\mu\nu}(\pm1)\,&\equiv&\,\sqrt{\frac{1}{2}}\left[\epsilon(\pm1)^{\mu}\epsilon(0)^{\nu}\,
+\,\epsilon(0)^{\mu}\epsilon(\pm1)^{\nu}\right],\nonumber\\
\epsilon^{\mu\nu}(0)\,&\equiv&\,\sqrt{\frac{1}{6}}\left[\epsilon(+1)^{\mu}\epsilon(-1)^{\nu}\,
+\,\epsilon(-1)^{\mu}\epsilon(+1)^{\nu}\right]\,+\,\sqrt{\frac{2}{3}}\epsilon(0)^{\mu}\epsilon(0)^{\nu}.
\end{eqnarray}
Due to the angular momentum conservation argument, the $\pm2$
polarizations do not contribute in these considered decays
\cite{wwprd83014008}. This can effectively simplify the following
perturbative calculations. What is more, the light-cone distribution
 amplitudes (LCDAs) of the tensor meson are antisymmetric under the
 interchange of momentum fractions of the quark and anti-quark in
 the SU(3) limit  \cite{zheng1zheng2}, due to the Bose statistics.
 \section{RESULTS AND DISCUSSION}
 The numerical results of these considered decay branching ratios and polarization fractions are displayed in Ref.\cite{zou1}.
 There are no direct CP asymmetries, because these decays do not have contributions from the penguin operators,
 and we find that compared with $B\rightarrow D^{(*)} T$
decays, the $B\rightarrow \bar{D}^{(*)}T$ decays are enhanced by the
CKM matrix elements $|V_{cb}/V_{ub}|^2$.
\begin{table}[t]
\begin{center}
\caption{Branching ratios (unit:$10^{-7}$) and the percentage of transverse polarizations $R_{T}$(unit:$\%$) of $B_{(s)}\rightarrow D^{*}T$ decays
calculated in the PQCD approach
 together with results from ISGW II model \cite{arxiv1010.3077,prd67014011}.}
\begin{tabular}{|l|c|c|c|c|c|}
\hline Decay Modes & Class & Br(PQCD) & Br(SDV \cite{arxiv1010.3077})& Br(KLO \cite{prd67014011})& $R_{T}$
\\
\hline
$B^{0}\rightarrow D^{*0}a_{2}^{0}$ & C & $1.34_{-0.53\,-0.19\,-0.15}^{+0.64\,+0.25\,+0.17}$& 0.50& ...& $47_{-4.5\,-1.6}^{+3.7\,+1.4}$\\
\hline
$B^{0}\rightarrow D^{*0}f_{2}$ & C & $2.70_{-1.02\,-0.30\,-0.33}^{+1.22\,+0.43\,+0.36}$ & 0.53&...&$26_{-4.0\,-1.1}^{+3.7\,+1.3}$ \\
\hline
 $B^{0}\rightarrow D^{*0}f_{2}^{\prime}$ & C&$0.052_{-0.02\,-0.005\,-0.006}^{+0.023\,+0.008\,+0.007}$ & 0.01 & ...&$26_{-4.0\,-1.1}^{+3.7\,+1.3}$ \\
\hline
$B^{0}\rightarrow D^{*0}K_{2}^{*0}$ & C&$60.5_{-21.3\,-9.15\,-7.56}^{+25.3\,+10.6\,+8.30}$ & 19 & 18 &$22_{-3.2\,-0.5}^{+2.7\,+1.0}$\\
\hline
\end{tabular}
\label{t1}
\end{center}
\end{table}

In table ~\ref{t1}, one can see that for the color suppressed (C)
decay modes, the predicted branching ratios in the PQCD approach are
larger than those of Ref.\cite{arxiv1010.3077} and
Ref.\cite{prd67014011}. For example, $B^{0}\rightarrow
\bar{D}^{0}f_{2}$, our predicted branching ratio is $\mathcal
{B}(B^{0}\rightarrow \bar{D}^{0}f_{2})\,=\,9.46\times10^{-5}$. It is
larger than other approaches, but agrees better with
the experimental data $(12\pm4)\times10^{-5}$ \cite{jpg37075021}.
For these color suppressed decay channels, the non-factorizable
contributions play dominant role in the amplitude, because the
facotrizable contribution is suppressed by the Wilson coefficient
$a_{2}$ $(C_{1}+C_{2}/3)\simeq0.1$, while the Wilson coefficient for
nonfactorizable contribution is $C_{2}\simeq1.0$. What is more, when
the emitted meson is the $D(\bar{D})$ meson or tensor meson, the
contributions for two nonfactorizable emission diagrams no longer
cancel with each other, because of the big difference between the
$c$ quark and the light quark in $D$ meson and the antisymmetry of
the tensor meson wave function.

\begin{table}[t]
\begin{center}
\caption{Branching ratios (unit:$10^{-7}$) and the percentage of transverse polarizations $R_{T}$(unit:$\%$) of $B_{(s)}\rightarrow D^{*}T$ decays calculated in the PQCD approach
 together with results from ISGW II model \cite{arxiv1010.3077,prd67014011}.}
\begin{tabular}{|l|c|c|c|c|c|}
\hline Decay Modes & Class & Br(PQCD) & Br(SDV \cite{arxiv1010.3077})& Br(KLO \cite{prd67014011})& $R_{T}$
\\
\hline
$B^{+}\rightarrow D^{*+}f_{2}^{\prime}$ & T & $0.29_{-0.11\,-0.04\,-0.02}^{+0.15\,+0.03\,+0.02}$& 0.21& ...& $25_{-1.0\,-1.8}^{+1.3\,+1.8}$\\
\hline
$B_{s}\rightarrow D^{*+}K_{2}^{*-}$ & T & $14.8_{-5.93\,-0.85\,-1.77}^{+7.42\,+0.90\,+1.94}$ & 12&...&$26_{-1.0\,-0.2}^{+1.2\,-0.1}$ \\
\hline
\end{tabular}
\label{t2}
\end{center}
\end{table}

From table~\ref{t2}, we can find that for those color allowed (T)
decay channels, our predicted branching ratios basically agree with
that predicted in naive factorization approach in
Ref.\cite{arxiv1010.3077}. The dominant difference is mainly caused
by parameter changes and the additional contributions from
nonfactorizable and annihilation diagrams. For some color allowed
decays, for example, $B^{0}\rightarrow D^{-}a_{2}^{+}$, there are no
contributions from factorizable emission diagrams, because the
emitted meson is the tensor meson $a_{2}^{+}$. For these decays, we
give the predictions for the first time by calculating the
contributions from non-factorizable diagrams and the annihilation
type diagrams.

\begin{table}[t]
\begin{center}
\caption{Branching ratios (unit:$10^{-5}$) and the percentage of transverse polarizations $R_{T}$(unit:$\%$) of $B_{(s)}\rightarrow \bar{D}^{*}T$ decays calculated in the PQCD approach
 together with results from ISGW II model \cite{arxiv1010.3077,prd67014011}.}
\begin{tabular}{|l|c|c|c|c|c|}
\hline Decay Modes & Class & Br(PQCD) & Br(SDV \cite{arxiv1010.3077})& Br(KLO \cite{prd67014011})& $R_{T}$
\\
\hline
$B^{0}\rightarrow \bar{D}^{*0}a_{2}^{0}$ & C & $39.3_{-11.1\,-0.50\,-1.34}^{+13.6\,+2.05\,+2.15}$& 12&7.8& $73_{-4.3\,-8.1}^{+4.6\,+9.0}$\\
\hline
$B^{0}\rightarrow \bar{D}^{*0}f_{2}$ & C & $38.2_{-11.6\,-1.22\,-1.30}^{+13.9\,+1.97\,+2.10}$ & 13&8.4&$70_{-5.6\,-6.3}^{+5.9\,+9.4}$ \\
\hline
 $B^{0}\rightarrow \bar{D}^{*0}f_{2}^{\prime}$ & C&$0.72_{-0.22\,-0.03\,-0.03}^{+0.26\,+0.02\,+0.04}$ & 0.26 & 0.11&$70_{-5.6\,-6.3}^{+5.9\,+9.4}$ \\
\hline
$B^{0}\rightarrow \bar{D}^{*0}K_{2}^{*0}$ & C&$5.32_{-1.42\,-0.66\,-0.22}^{+1.69\,+0.79\,+0.32}$ & 1.3 & 1.1 &$71_{-1.6\,-8.8}^{+1.8\,+8.6}$\\
\hline
\end{tabular}
\label{t3}
\end{center}
\end{table}

According to the power counting rules in the factorization
assumption, from the quark helicity analysis, the longitudinal
polarization should be dominant \cite{helicity1,helicity2}. However,
 in table~\ref{t3}, for those color suppressed (C) $B\rightarrow
\bar{D}^{*}T$ decays with the $\bar{D}^{*}$ emitted, the transverse
polarization fractions are about $70\%$. For these decays, the
$\bar{c}$ quark and the $u$ quark in $\bar{D}^{*}$ meson are all
produced through $(V-A)$ current, and the $\bar{c}$ is right-handed;
while the $u$ quark is left-handed. As a result, the $\bar{D}^{*}$ meson
is longitudinally polarized. We know that the helicity of massive
quark can flip easily. So the $\bar c$ quark can flip easily from
right handed to left handed, then the polarization of the
$\bar{D}^{*}$ meson can be equal to -1. On the other hand, because
of the additional contribution of orbital angular momentum, the
recoiled tensor meson can also be transversely polarized with
polarization $\lambda=-1$. Thus, the transversely polarized
contributions are no longer small. In table~\ref{t1}, for color
suppressed (C) $B\rightarrow {D}^{*}T$ decays with $D^{*}$ meson
emitted, the percentage of transverse polarizations are only at the
range of $20\%$ to $30\%$. After the similar analysis, we know that
the emitted $D^{*}$ meson can also be transversely polarized with
the polarization $\lambda=+1$. According to the angular momentum
conservation, the recoiled tensor meson must be also transversely
polarized with polarization $\lambda=+1$. In this case, the tensor
meson needs contributions from both orbital angular momentum and
spin, so the situation is symmetric. As we know that the wave
function of tensor meson is asymmetric. Therefore the transversely
polarized contribution is suppressed, because of
 Bose statistics.

 \begin{table}[t]
\begin{center}
\caption{Branching ratios (unit:$10^{-7}$) and the percentage of transverse polarizations $R_{T}$(unit:$\%$) of $B_{(s)}\rightarrow D^{*}T$ decays calculated in the PQCD approach
 together with results from ISGW II model \cite{arxiv1010.3077,prd67014011}.}
\begin{tabular}{|l|c|c|c|c|c|}
\hline Decay Modes & Class & Br(PQCD) & Br(SDV \cite{arxiv1010.3077})& Br(KLO \cite{prd67014011})& $R_{T}$
\\
\hline
$B^{+}\rightarrow D^{*+}K_{2}^{*0}$ & A & $18.2_{-5.15\,-2.15\,-2.70}^{+4.77\,+0.21\,+2.00}$& ...&...& $82_{-2.9\,-2.7}^{+2.1\,+3.8}$\\
\hline
$B^{+}\rightarrow D_{s}^{*+}f_{2}^{\prime}$ & A & $21.6_{-6.03\,-2.32\,-2.70}^{+6.77\,+1.00\,+3.00}$ & 4.0&2.0&$83_{-5.3\,-1.9}^{+5.2\,+1.9}$ \\
\hline
 $B^{+}\rightarrow D_{s}^{*+}\bar{K}_{2}^{*0}$ & A &$1.25_{-0.34\,-0.16\,-0.15}^{+0.36\,+0.06\,+0.16}$ &... & ...&$81_{-1.8\,-3.3}^{+1.6\,+3.7}$ \\
\hline
\end{tabular}
\label{t4}
\end{center}
\end{table}

 We also find very large transverse polarizations up to 80\% for the
W annihilation (A) type  $B\rightarrow D^{*}T$ decays in table
~\ref{t4}. The light quark and unti-quark produced
through hard gluon are left-handed or right-handed with equal
opportunity. So the $D^{*}$ meson can be longitudinally polarized,
and also be transversely polarized with polarization $\lambda=-1$.
For the tensor meson, the anti-quark from weak interaction is
right-handed; while the quark produced from hard gluon can be either
left-handed or right-handed. When taking into account the additional
contribution from the orbital angular momentum, the tensor meson can
be longitudinally polarized or transversely polarized with
polarization $\lambda=-1$. So the transverse polarization can become
so large with additional interference from other diagrams.
\section{SUMMARY}
We investigate the $B_{(s)}\rightarrow D^{(*)}T,\bar{D}^{(*)}T$
decays within the framework of perturbative QCD approach. We find
that the nonfactorizable and annihilation type diagrams play
important roles in the amplitude, especially for those color
suppressed channels and the decays with a tensor meson emitted. For
the decays with a tensor meson emitted, we give the predictions
for the first time. For those color suppressed $B_{(s)}\rightarrow
\bar{D}^{*}T$ decays, the nonfactorizable diagrams provide sizable
transversely polarized contributions. For those W annihilation type
$B\rightarrow D^{*}T$ decays, the factorizable annihilation diagrams
give large transverse polarized contributions up to $80\%$.
%


\begin{acknowledgments}
 This Work is partly supported by
the National Science Foundation of China under the Grant
No.11075168, 11228512 and 11235005. We thank  Xin Liu and
 Run-Hui Li for helpful discussions.
\end{acknowledgments}

\end{document}